\documentclass{article}

\usepackage{PRIMEarxiv}

\usepackage[utf8]{inputenc} 
\usepackage[T1]{fontenc}    
\usepackage{hyperref}       
\usepackage{url}            
\usepackage{booktabs}       
\usepackage{amsfonts}       
\usepackage{nicefrac}       
\usepackage{microtype}      
\usepackage{lipsum}
\usepackage{fancyhdr}       
\usepackage{graphicx}       
\graphicspath{{media/}}     
\usepackage{comment}
\usepackage{amsmath}
\usepackage{subfig}
\usepackage{wrapfig}
\usepackage{color}
\title{Classical ensemble of Quantum-classical ML algorithms for Phishing detection in Ethereum transaction networks
\thanks{\textit{\underline{Citation}}: 
\textbf{Authors. Title. Pages.... DOI:000000/11111.}} 
}

\author{
  Anupama Ray\\
  IBM Research India\\
  \texttt{anupamar@in.ibm.com}\\
   \And
   Sai Sakunthala\\
   IIT Madras\\
   \texttt{saisakunthala1204@gmail.com}\\
    \And
    Vishnu Ajith\\
    IIIT Kottayam\\
    \texttt{vishnuajith@gmail.com}\\
   \And
  Dhinakaran Vinayagamurthy \\
  IBM Research India\\
  \texttt{dvinaya1@in.ibm.com} \\
}

\begin{document}
\maketitle





\begin{abstract}
Ethereum is one of the most valuable blockchain networks in terms of the total monetary value locked in it, and arguably been the most active network where new blockchain innovations in research and applications are demonstrated. But, this also leads to Ethereum network being susceptible to a wide variety of threats and attacks in an attempt to gain unreasonable advantage or to undermine the value of the users. Even with the state-of-art classical ML algorithms, detecting such attacks is still hard. This motivated us to build a hybrid system of quantum-classical algorithms that improves phishing detection in financial transaction networks. This paper presents a classical ensemble pipeline of classical and quantum algorithms and a detailed study benchmarking existing Quantum Machine Learning (QML) algorithms such as Quantum Support Vector Machine (QSVM) and Variational Quantum Classifier (VQC). With the current generation of quantum hardware available, smaller datasets are more suited to the QML models and most research restricts to hundreds of samples. However, we experimented on different data sizes and report results with a test data of 12K transaction nodes, which is to the best of the authors knowledge the largest QML experiment run so far on any real quantum hardware. The classical ensembles of quantum-classical models improved the macro F-score as well as phishing F-score. One key observation is QSVM constantly gives lower false positives, thereby higher precision compared with any other classical or quantum network, which is always preferred for any anomaly detection problem. This is true for QSVMs when used individually or via bagging of same models or in combination with other classical/quantum models making it the most advantageous quantum algorithm so far. The proposed ensemble framework is generic and can be applied for any classification task. All codes are available as tutorial notebooks: \url{https://github.com/anupamaray/EnsembleQML_application} along with the data required to reproduce the results reported.

\end{abstract}

\section{Introduction}
\label{sec:intro}

Quantum Machine Learning (QML) is one of the most promising directions where quantum computing principles are expected to make an impact in the near future \cite{Seth}. 
This is attributed to the interplay between ideas from both machine learning and quantum computing that can help each other. For instance, quantum kernels can improve ML algorithms and (classical) ML can help in better error mitigation and compilation for quantum hardware. 
For certain problems, it has been established that quantum machine learning algorithms, even with classical access to data, can provably outperform classical algorithms \cite{Liu_2021}. Despite the promise, the existing QML algorithms do not far better than classical ML algorithms on most real world datasets, and in many cases, perform much worse. Traditionally, ML research has used an ``ensemble'' of multiple learners such that the ensemble performs better than any of the individual models. Different techniques like bagging, boosting and stacking had been proposed for creating ensembles. Recently, the QML community has been taking inspiration from these and propose ideas for quantum ensembles. 
\cite{Maria} proposed a quantum equivalent of Bayesian model averaging, but it was later shown by \cite{Mariadequantize} that their algorithm can be efficiently translated into a classical algorithm and thus obviating any potential quantum speedups.
Quantum Boosting \cite{QBoost} on the other hand assumes to execute Quantum Phase Estimation algorithm which cannot be realized in the noisy quantum computers in the near term. \cite{Macaluso1} proposed an idea based on bagging which was extended in \cite{Macaluso2} where they provide a framework to generate different transformations of training set and a quantum classifier is applied to obtain classifications in superposition.  

In this paper, we propose a classical ensemble of quantum and classical ML algorithms where we are able to achieve better results by stacking different quantum and classical algorithms and training with their predictions. We experiment on improving phishing detection in the Ethereum blockchain network. Ethereum blockchain and smart contract architecture has gained prominence, and most innovative decentralized applications are built to be compatible with the Ethereum network. Hence, the phishing activities have become an urgent threat to the security of the blockchain system. Although there have been studies on generic anomaly detection, research on anomaly detection in blockchain networks has been limited mainly due to three reasons: a lack of auditing, largely unlabeled data with space-time constraints when studying as a graph problem and extreme high class imbalance between non-phishing and phishing transactions.

In order to understand the maximum accuracy achievable classically, we apply several state-of-art classical algorithms on Ethereum data. We then proceed to perform exhaustive experimentation using different quantum algorithms such as Variational Quantum Classifiers (VQC), and Quantum Support Vector Machines(QSVM). Since VQCs are parameterized quantum circuits similar to neural networks, we tried several ansatz (parameterized circuits), layering of ansatz, and different types of feature maps. In QSVMs, we tried two different implementations of QSVM - one that estimates quantum kernels in a gate-based quantum computer, and another that formulates SVM as a Quadratic Unconstrained Binary Optimization (QUBO) to use quantum annealing based computers.

Since performance of VQCs is completely dependent on the choice of ansatz and currently all research works use heuristics for ansatz selection, we delve deeper into ansatz quality metrics and its relation with precision and recall. We compute expressibility and entangling capacity of each quantum circuit as proposed in \cite{Sim_2019}. We perform correlation studies between these ansatz metrics and precision-recall-fscore to understand the amount of expressibility or entangling capacity needed in the problem, which helps chose ansatz based on metrics instead of heuristics. Another challenge of currently available quantum algorithms for near-term quantum computers is limited training data size, thus most works are restricted to experiments using data in the scale of few hundreds \cite{Frerdrik}. In this work, we experimented on scaling the training data of quantum algorithms from few hundreds to few thousands and analyze the data capacity of quantum models, their learning ability as well as training time. Finally, after a detailed analysis of each model, we create a framework of ensembling quantum-classical algorithms to improve phishing detection and make use of complimentary nature of classical and quantum algorithms.

To summarize, the main contributions of this work are as follows:
\begin{itemize}
\item Creation of a end-to-end trainable framework for a classical ensemble of quantum ML and classical ML models using stacking and bagging techniques.
\item Benchmarking QML algorithms (VQC and QSVM) with exhaustive experimentation using different feature maps (first order Pauli Z evolution, second order Pauli Z evolution and amplitude encoding). In VQC, we use 19 benchmarked ansatz and multiple layers of ansatzes. In QSVM, we experimented with two different implementations one using gate-base quantum computers and the other using annealers.
\item Exhaustive experiments on IBM Quantum Hardware (7Qubit and 27Qubit systems) and DWave Annealer 
\item Study of correlation between the ansatz quality metrics such as expressibility and entangling capacity with the performance metrics of phishing detection (precision-recall-fscore).
\item Phishing detection in Ethereum network using quantum algorithms, which to the best of our knowledge has not been tried before. 
\end{itemize}

The methods proposed and developed in this work are generic and can be applied to any classification task. All the configurations for each experiment were performed on IBM statevector simulator and we report mean of 5 runs for each experiment. The best configurations are selected along with optimal hyperparameters to run on actual IBM Quantum Hardware.

\section{Related Work}
\label{sec:related}
The research of anomaly detection using quantum algorithms is nascent.
On Credit card fraud detection, \cite{ZLW21} presented a Variation Quantum Boltzmann Machine based method, \cite{HOR21} developed a variational quantum-classical Wasserstein GAN, and \cite{KM22} study the use of one-class SVM in an unsupervised setting. Classically however, there have been extensive research on phishing detection, or more generally, anomaly detection  \cite{MWX+21, PSCH21,KIJ13}. For phishing detection in Ethereum networks, the problem of interest, the state-of-art classical solution use the Graph Convolutional Neural Networks (GCN) \cite{ZTXM18}. Two concurrent papers \cite{CGCZL20, EtherACMToIT} trained GCN on a labeled version of a snapshot of the transaction graph of the Ethereum network obtained from etherscan.io. While in \cite{EtherACMToIT} authors present an array of 200 statistical features that could be important for phishing detection, \cite{CGCZL20} uses 8 financial features out of the 200 features proposed and use GCN to obtain the classical state-of-art on this problem. 


More generally, there are a variety of machine learning tasks that quantum computers promise an improvement over classical computers \cite{BLSF19, BCK+22}. The work on classification problems can be broadly categorized into those based on variational models \cite{MNKF18, HCT+19, SBSW20, PCGL20, NCW21} and kernel-based methods \cite{HCT+19, SK19, Sch21}. There has also been research on identifying anomalous quantum data using quantum algorithms \cite{LR18, KMFB21}, but this line of work is not relevant for our work as we deal with a classical dataset. There has also been some progress made on realizing quantum convolutional neural networks \cite{CCL19, CWZ+22}, and quantum graph neural networks \cite{VML+19}. The work on other supervised QML tasks are on regression \cite{MNKF18, PEK22}, solving linear \cite{XSE+21, CSWW19} and differential equations \cite{LJM+20, KPE21, PEK22}. Even more generally, quantum computing has shown promises of speedups in financial applications \cite{EGM+20} including portfolio optimization \cite{KPS19, YMHP21}, risk prediction \cite{WE19, SMWZ22} and pricing of financial contracts \cite{SES+20, CKM+21}.


\section{Quantum Machine Learning Models: Background}
\label{sec:Background}

In this section, we describe the two quantum algorithms - QSVM and VQC that have been used in this paper. For any quantum model to work on classical data, the first step is to be able to load the classical data as quantum states. This step is called \textbf{data encoding or mapping}, and we use parameterized quantum circuits as feature maps. In this paper we have used three types of encoding schemes which are described below. Assuming $n$ training data points and $m$ features of every data point, the classical training data can be written as
\begin{align}
    &X=\{x_1,\dots,x_n\},where\;n=no.\;of\;data\;points,\\
    &x_i=\{x_i^{(1)},\dots,x_i^{(m)}\},where\;m=no.\;of\;features
\end{align}
In order to load these classical features as quantum states we can apply different types of transformations, such that the qubits of a quantum circuit now represents each datapoint and its features.

\subsection{Amplitude Encoding}
In the amplitude encoding, data is encoded into the amplitudes of a quantum state. Consider a data point vector $x_i$ with m features and we normalize it to
\begin{align}
    &x_{norm}=\frac{1}{\Sigma_j(x_i^{(j)})^2}\{x_i^{(1)},\dots,x_i^{(m)}\}
\end{align}
Now to encode m features we require $\lceil log_2m \rceil$ number of qubits. Then the corresponding dataset in computational basis state can be written as
\begin{align}
    let\;p=\lceil log_2m \rceil,\;\;&\vert i\rangle=p\;qubit\;state\\
    \vert \psi_x\rangle=\sum_i(x_{norm})_i\vert i\rangle,&\;\;i\in\{1,\dots,m\}
\end{align}
the state $\vert \psi_x\rangle$ is our final amplitude encoded state of m feature vector using p qubits. Since we had identified 7 financial features, we padded the feature set with a vector of zeros to match the cardinality while encoding and were able to encode it in just 3 qubits instead of 8. 

\subsection{Z feature map}
Zfeature map is a quantum circuit composed of only single qubit operations that evolves the state of each qubit independently by applying Hadamard gate and parameterized unitary gates. The resulting circuit contains no interactions between features of the encoded data, and therefore no entanglement. Here the number of qubits required are same as number of features. For a data point vector $x_i$ with m features $x_i=\{x_i^{(1)},\dots,x_i^{(m)}\}$
\begin{align}
    &Z feature map = U_{\phi(x)}H^{\otimes m}\\
    &where\;U_{\phi(x)} = exp\left(j\sum_{k=1}^mx_i^{(k)}Z_k\right),\\
    &Z=pauli\;Z\;gate,\;\;j = imaginary\;unit\nonumber
\end{align}

\subsection{ZZ featuremap}
ZZ feature map involves entanglement between qubits along with single qubit operations. The number of qubits required are same as number of features. For a vector $x_i$ with m features $x_i=\{x_i^{(1)},\dots,x_i^{(m)}\}$
\begin{align}
    ZZ feature& map = U_{\phi(x)}H^{\otimes m}\\
    where\;U_{\phi(x)} = &exp\left(j\sum_{k\in [m]}\phi_k(x_i)\prod_{l\in k}Z_l\right), \\
    [m]=\{1,\dots,m,&(1,2),(1,3),\dots,(m-1,m)\}\nonumber\\
    where\;&\phi_p(x_i) = x_i^{(p)}\;and\\
    \;\phi_{(p,q)}(x_i)=&(\pi-x_i^{(p)})(\pi-x_i^{(q)})\nonumber\\
    Z=pauli\;Z&\;gate,\;\;j = imaginary\;unit\nonumber
\end{align}

Efficient utilization of qubits in data encoding is still an open research problem and there are several other data encoding schemes, which are beyond the scope of this paper. Once we have loaded the data, we have different flows for QSVM and VQC which is explained in the following subsections.

\subsection{Variational Quantum Classifier (VQC)}
\begin{figure}
    \centering
    \includegraphics[scale = 0.50]{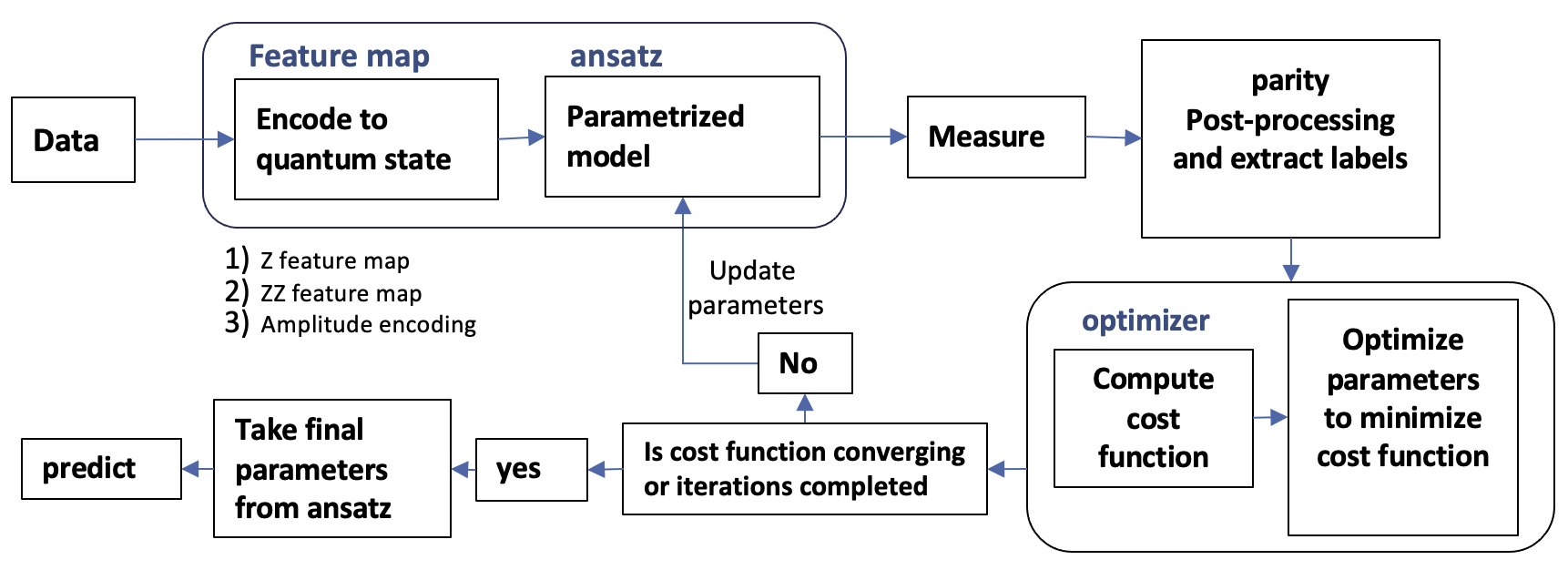}
    \caption{Implementation of variational quantum classifier}
    \label{vqc}
\end{figure}
Figure \ref{vqc} shows the flow of each step in order to build a VQC model. First we load the classical data into quantum state using the above mentioned encoding schemes. In a Variational Quantum Classifier, we create layers of learnable parameterized quantum circuits (also called ansatz) after the feature mapping step, which are measured and then after applying some postprocessing to get the desired output passed through a cost function. We update the parameters of the ansatz to minimize the overall cost function, much like training weights of a neural network. In the current implementation, the measurement results were interpreted based on the parity of the measurement outputs, where even parity is considered as label +1 and odd parity as -1. After obtaining labels from parity post-processing, the classical optimizer calculates cost function, and optimizes the parameters of ansatz, till the classical optimization iterations complete or till the cost function converges. For inferencing too, we use multiple shots and the most probable label is selected as final label for each test data. We used Constrained Optimisation By Linear Approximation (COBYLA) as the classical optimizer, which is a gradient-free optimizer to learn the ansatz parameters. There are several other classical optimizers (gradient-based and gradient-free) that can also be applied to such networks, however we present results with COBYLA only in this paper. Amplitude encoding does not support gradient-based encoding schemes, thus for uniformity we applied the same optimizer for all ablation studies. Initial tests, hyperparameter optimizations and experiments with different data sizes are run on a noiseless state vector simulator provided by IBM Quantum. The best performing ansatzes with other chosen hyperparameters are used to conduct experiments on real IBM Quantum hardware. 

\subsection{QSVM using annealing (QSVM\_Dwave)}
Quantum annealing is a heuristic approach to adiabatic quantum computing and claims advantage over classical annealing algorithms because of quantum tunneling through local optima unlike the classical counterpart where optima would have to be found by thermal fluctuations. Quantum annealers are composed of superconducting qubits and couplers that form a time dependent hamiltonian which can be tuned to a Quadratic Unconstrained Binary Optimization (QUBO) formulation.

In this work, we use classical SVM model for QUBO formulation by considering it as an optimization model. Given training data matrix $X \in R^{N \times d}$ and training labels $Y \in \{-1,+1\}^N$, where $N$ is the number of training data points, the hyperplane is determined by weights, $w\in R^d$, and bias, $b\in R$, that separates the training data into binary classes. Mathematically,
\begin{align}
    \min_{w,b}&\;\frac{1}{2}\|w\|^2\\
    subject\;to\; \; y_i(w^Tx_i+b)\geq 1,  \;\;\;\forall &i=1,2,\dots,N \nonumber \;\;\;\forall x_i\in rows\;in\;X \;\;\;\forall y_i\in Y
\end{align}
The lagrange of the optimization model in primal form is formulated
\begin{align}
 \mathcal{L}(w,b,\lambda)=\frac{1}{2}\|w\|^2-\sum^{N}_{i=1}\lambda_i[y_i(w^Tx_i+b)-1]
\end{align}
where, $\lambda$ is the vector containing all the Lagrangian multipliers, i.e. $\lambda=[\lambda_1\dots \lambda_N]^T, \; with\; \lambda_i\geq 0,\; \forall i$. Each Lagrange multiplier corresponds to one training data point and represents significance of that particular data point in determining hyper-plane. Converting the above primal problem to its dual form yields the following QUBO
\begin{align}
\label{13}
\min_{\lambda}\mathcal{L}(\lambda)=\frac{1}{2}\sum^{N}_{i=1}\sum^{N}_{j=1}\lambda_i\lambda_j y_iy_j(x_i^Tx_j)-\sum^{N}_{i=1}\lambda_i
\end{align}
and $\lambda_i,\lambda_j\geq0$,  $\forall i,j$. We use a kernel function to project the input data to higher dimensions and then use SVM on the higher dimensional data. The kernel matrix for rbf kernel is defined as
\begin{align}
K_{ij}=e^{-\lVert x_i-x_j \rVert^2/2\sigma^2},\;\;\forall i,j
\end{align}
We are using rbf kernel with $\sigma$=150, which we found by trail and error. Substituting kernel in equation \ref{13} yields the final QUBO equation:
\begin{align}
\min_{\lambda}\mathcal{L}(\lambda)=\frac{1}{2}\sum^{N}_{i=1}\sum^{N}_{j=1}\lambda_i\lambda_jy_iy_j(K_{ij})-\sum^{N}_{i=1}\lambda_i
\end{align}

The dual form of the Lagrange of the SVM formulation in matrix form is as follows:
\begin{align}\label{eqn:dual}
    \min_{\lambda}\mathcal{L}(\lambda)=\frac{1}{2}\lambda^T(K\odot YY^T)\lambda-\lambda^T 1_N, \; \lambda\geq 0_N
\end{align}
where K is kernel matrix, $1_N$ and $0_N$ represent N-dimensional vectors of ones and zeros respectively, $\odot$ is the element-wise multiplication operation. 
The QUBO matrix is given as input to a quantum annealer\footnote{https://cloud.dwavesys.com/leap/signup/}, which solves the minimization objectives and returns the Lagrange multipliers (binary), which are the support vectors. 

Precision vector is introduced to have integer support vectors instead of only binary and the dimension of precision vector depends on the range of integer values for support vector. Precision vector has powers of 2 as elements, and here we use precision vector $p = [2^0,2^1]$ to get the final QUBO matrix. Let $\hat{\lambda}=[\lambda_{11},\lambda_{12},\dots,\lambda_{N1},\lambda_{N2}]$, we pass the QUBO matrix to an annealing quantum computer. The ground state of the time-dependent Hamiltonian is composed of the $\hat{\lambda}$ values, that minimize the formulation. 
\begin{align}
    \min_{\hat{\lambda}}\mathcal{L}(\hat{\lambda})=\frac{1}{2}\hat{\lambda}^TP^T(K&\odot YY^T)P\hat{\lambda}-\hat{\lambda}^Tp^T1_N
\end{align}
where $P = I_n\otimes p$ and $\lambda=P\hat{\lambda}$. The annealer returns the $\hat{\lambda}$, which we used to calculate Lagrange multipliers using precision vector, and $\lambda$ is our final support vector values. Prediction for unseen data is done by the equation \ref{eqn:pre} after determining $\lambda$ by solving the QUBO.
\begin{align}\label{eqn:pre}
    \mathsf{label}(x)=\mathrm{sign}&\left(\sum^{N}_{i=1} \lambda_iy_i(K_{xi})+b\right)\\
    b = \mathsf{mean}(y_i-w^Tx_i),&\;\;\;\text{where} \;i\in[0,\dots,N],\\
    w^Tx_i=\sum^{N}_{j=1}&\lambda_jy_jK_{ji}\nonumber
\end{align}

\subsection{QSVM using quantum circuits (QSVM\_qiskit)}
A quantum kernel is the inner product of quantum feature maps and the main idea of QSVM is that if we choose a quantum feature map that is not easy to simulate with a classical computer there is scope for a quantum advantage  \cite{HCT+19}. 
Quantum kernels are more efficient in projecting the data to higher dimensions. Generally classical kernels can be represented as $K(x_i,x_j)=K_{ij}=f(x_i,x_j)$, where $f$ is a function used to project data. Instead of functions, we use quantum feature maps to project classical data into quantum states and then use them to form a kernel matrix used to project data points. 
First, we initialize circuit in $\vert 0\rangle$ state, then we apply feature maps $\phi(\Vec{x_i})$ and $\phi^\dagger(\Vec{x_j})$ to the circuit, which changes the state to $\phi^\dagger(\Vec{x_j})\phi(\Vec{x_i})\vert 0\rangle$. We now measure the circuit in Z bases and the kernel element is obtained by taking the probability of measuring $\vert 0\rangle$. Mathematically, kernel elements can be represented as
\begin{align}
    K(\vec{x_i},\vec{x_j}) =K_{ij}=& \mathrm{Pr}[\mathsf{measure}\;\vert 0\rangle]\\
    K(\vec{x_i},\vec{x_j}) =K_{ij}=& \vert\langle 0\vert \phi^\dagger(\vec{x_j})\phi(\vec{x_i})\vert 0\rangle\vert^2\\
    K(\vec{x_i},\vec{x_j}) =K_{ij}=& \vert\langle\phi^\dagger(\vec{x_j})\vert\phi(\vec{x_i})\rangle\vert^2
\end{align}
We tried three different feature maps namely Z feature amp, ZZ feature map and amplitude encoding, and used the kernel matrix in dual form of SVM as in Equation~\ref{eqn:dual} and performed optimization same as classical SVM and predict data using Equation~\ref{eqn:pre}.

 Quantum kernels are more efficient in projecting the data to higher dimensions. Generally classical kernels can be represented as $K(x_i,x_j)=K_{ij}=f(x_i,x_j)$, where $f$ is a function used to project data. Instead of functions we use quantum feature maps to project classical data into quantum states, and then use them to form kernel matrix, used to project data points. 
 The circuit representation of kernel is shown in figure\ref{kernel}, where $\phi(\Vec{x})$ is the feature map used in kernel
 First we initialize circuit in $\vert 0\rangle$ state, then we add feature maps $\phi(\Vec{x_i})$ and $\phi^\dagger(\Vec{x_j})$ to the circuit, which changes the state to $\phi^\dagger(\Vec{x_j})\phi(\Vec{x_i})\vert 0\rangle$, now we measure the circuit in Z bases, the kernel element is obtained by taking the probability of measuring $\vert 0\rangle$ , mathematically kernel elements can be represented as
\begin{align}
     K(\vec{x_i},\vec{x_j}) =K_{ij}=&Probability[measure\;\vert 0\rangle]\\
     K(\vec{x_i},\vec{x_j}) =K_{ij}=& \vert\langle 0\vert \phi^\dagger(\vec{x_j})\phi(\vec{x_i})\vert 0\rangle\vert^2\\
     K(\vec{x_i},\vec{x_j}) =K_{ij}=& \vert\langle\phi^\dagger(\vec{x_j})\vert\phi(\vec{x_i})\rangle\vert^2
 \end{align}
 We have tried three different feature maps namely Z feature amp, ZZ feature map and amplitude encoding, and used the kernel matrix in dual form of SVM as in Equation~\ref{eqn:} and performed optimization same as classical SVM and predict data using Equation~\ref{eqn:}.

\end{comment}

\section{Proposed Methodology}
\label{sec:Methodology}

In order to understand the issues faced by classical ML models on this problem, we experimented with several classical algorithms. Since classical algorithms do not suffer from any data limitations in their training, we used 80\% of the total three million data available for training and the remaining 20\% for testing. This experiment was done to understand the skyline performance on phishing detection possible today. Light Gradient Boosting Model (LGBM) outperformed other classical models in terms of recall, but suffered from very low precision. 
Since this is a graph data, using the full graph in a Graph Convolutional Neural Network (GCN) gave state-of-art results but took 17 hour of training ($\sim$8K epochs) to converge. On the other hand, LGBM completes training in less than 2 minutes under 10 epochs but shows comparable results with GCN only in terms of recall.

Near-term quantum algorithms such as VQC or QSVM are expected to be able to generate correlations that are hard to represent classically. In this work, we benchmark VQC using different ansatzes and different feature maps. Also for QSVM, we tried different formulations - namely the use of Quantum kernels with Qiskit and QUBO formualtion of QSVM using annealers to understand the capability of these models on the same problem.

\noindent\textbf{Ansatz quality metrics}. Akin to the selection of a feature map, the selection of the ansatz has been mostly through trial and error\footnote{An exception is in the domain of quantum chemistry, where they choose ansatz inspired from an extensive prior research on classical approximations of these problems.}. We attempt a systematic approach to ansatz selection by using the quality metrics (expressibility and entangling capacity) proposed in \cite{Sim_2019}. At a high level, expressibility of a n-qubit ansatz is measured by its capacity to explore the $2^n$ dimensional Hilbert space. We can calculate expressibility as the distance between a uniform distribution of quantum states and the distribution of quantum states obtained from a uniform sampling of the parameters in the ansatz. And, entangling capacity quantifies the ability of the ansatz to generate entangled states. In this work, we choose multiple ansatzes with varying expressibility and entangling capacity and understand the correlation between the metrics and the performance of the resulting model. We have used two ways to compute Entangling capacity - using Meyer Wallach measure (as proposed in the paper\cite{Sim_2019}) and Von-Neumann entropy measure. These correlation studies can help choose an ansatz for the variational algorithm in consideration for that application, in terms of how much entangling capacity is ideally required to improve results and so on.

\subsection{Classical Ensembles of Quantum-Classical ML Models}
In this work, we create classical ensembles of quantum and classical models to be able to leverage their individual advantages. Ensembling techniques like bagging and stacking have been implemented on all combinations of quantum and classical models and best results have been reported. 

\subsection*{Stacking}
\begin{figure}%
    \centering
    \includegraphics[width = 1\textwidth]{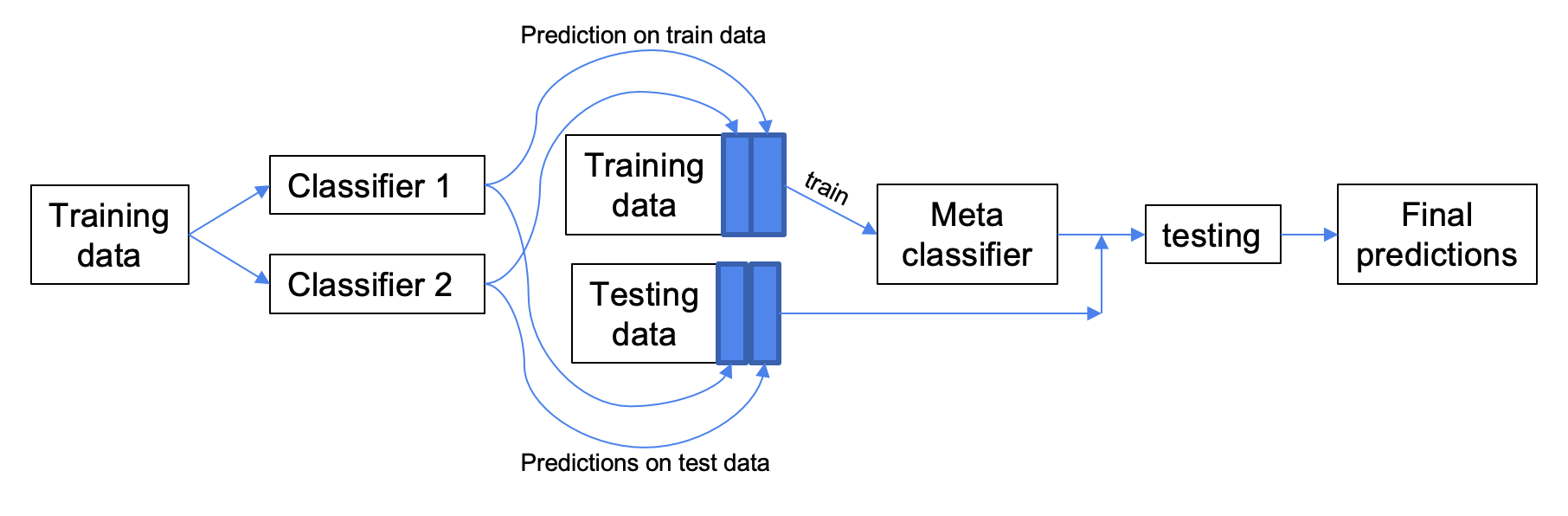}%
    \caption{Two level stacking technique implemented using various ML models}%
    \label{stacking}%
\end{figure}

We implemented two-level and three-level stacking of classifiers, a flow of two-level stacking is shown in Figure \ref{stacking}. In two level stacking, we have single or multiple base models and one meta classifier. After the data has been split into training and testing set, every base model is trained with the training data and tested over training data to obtain prediction vectors. Prediction vectors are appended as features to original training data to obtain the final training data for meta classifier. As the training and testing data should have same number of features, all the trained base models are tested over testing data set to obtain similar prediction vectors, and appended to testing data set to obtain final testing data for meta classifier. We performed this two level stacking with various base models and meta classifiers, with 160p-160np (p=phishing and np=non-phishing) as training data and 1000p-10000np as testing data. 

In three level stacking, we have level-0 classifiers, level-1 classifiers and one meta classifier. Similar to two level stacking, we trained the base models and the prediction vectors of models over training data are appended as features to original training data, and prediction vectors over testing data are appended to original testing data. Now, this new training data is used to train level-1 classifiers, the prediction vectors of training and testing data are again appended as features to obtain final training and testing data for meta classifier. The meta classifier is then trained and tested over this data to obtain final predictions. We performed this three level stacking with GCN as base model, QSVM\_qiskit and LGBM as level-1 classifiers and logistic regression as meta classifier, with the same train and test data.

\subsection*{Bagging}
\begin{figure}%
    \centering
    \includegraphics[width = 1\textwidth]{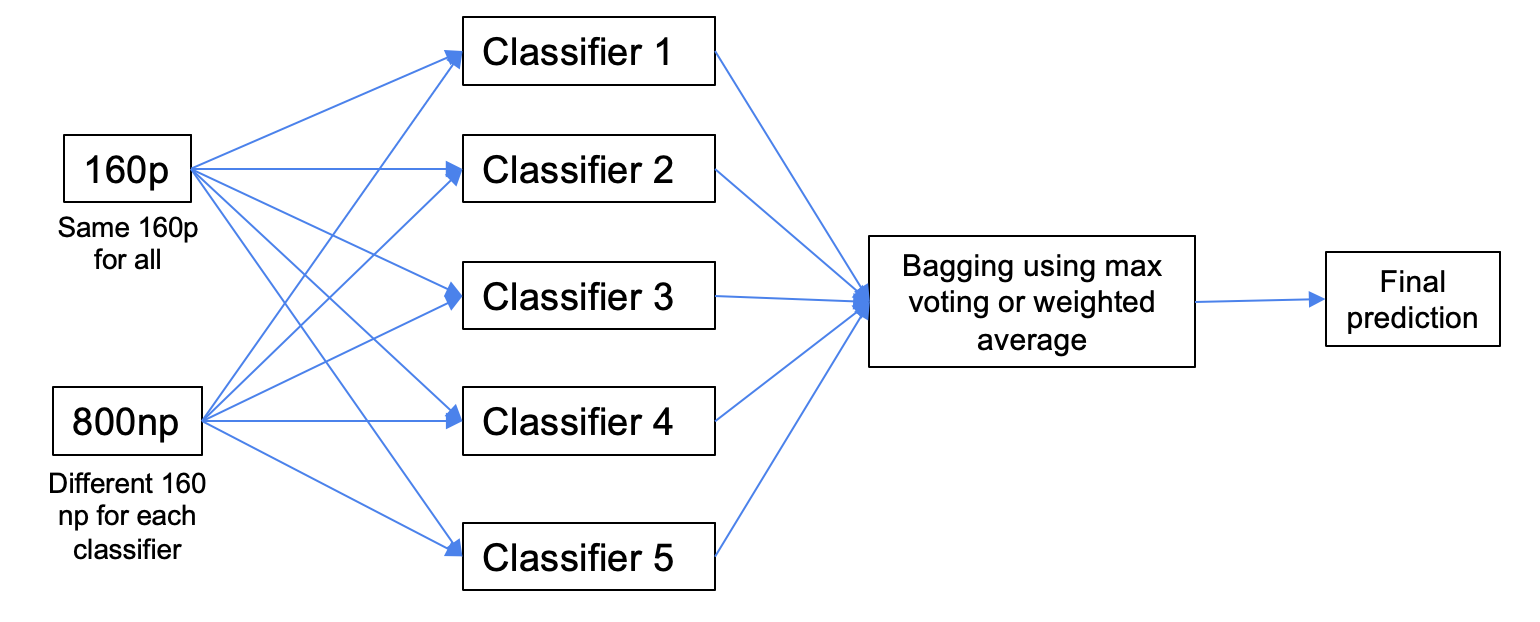}%
    \caption{Bagging technique implemented on three quantum models}%
    \label{bagging}%
\end{figure}

Bagging has been implemented on three quantum models by taking five models of same type each trained with 160p-160np data set. While the 160 phishing nodes remained same for all five models, we randomly sample different 160 non-phishing data points for each model. All such ensembles are tested on same test data (1000p-10000np) to obtain five prediction vectors. The predictions are combined by using maximum voting bagging and weighted average bagging taking phishing f1-scores as weights to obtain final prediction using each model. This is summarized in figure \ref{bagging}.

\section{Dataset and Feature Extraction}
\label{sec:dataset}

Ethereum is an open-source, public blockchain platform. Etherscan\footnote{{https://etherscan.io}} is a block explorer and analysis platform for Ethereum wherein some labeled data of phishing accounts can be obtained. All of these phishing nodes are derived from public reporting of phishing activities, as labeling samples otherwise is extremely costly and requires a lot of human effort. The number of phishing accounts reported is significantly smaller and the remaining unlabeled nodes are 2500 times more than the phishing nodes. Thus, there is a huge class-imbalance in the data, which becomes a major challenge for any prediction algorithm (classical or quantum). 

We crawled 3 million nodes which had 1165 phishing nodes (0.039\%) and 2972324 non-phishing nodes. \cite{EtherACMToIT} proposed 8 statistical features on the directed multigraph, of which we use 7, namely in-degree, out-degree, degree, in-strength, out-strength, strength and number of neighbors. In-degree refers to the number of incoming edges or transactions to the current node and out-degree is the number of outgoing transactions. Degree is the total number of edges or transactions for each node, and a lot of unevenness has been found in the in- or out-degree for phishing nodes \cite{EtherACMToIT}. Coming to the actual transactional value, we compute the in-strength as the total amount of incoming ether and out-strength as the total amount of ether sent to other nodes. The node's total transaction value is called strength of a node. Another factor discriminating between phishing and non-phishing nodes is that the total strength of phishing nodes is much higher than the non-phishing nodes. Apart from the directional and transactional features, we also compute the number of neighbors of each node. Since these are statistical features computed over a multigraph, the number of neighbors and the degree of a node may not be equal. For example, an account might have only one neighbor but multiple transactions between them would increase the value of degree for that node. Since phishing tends to increase the number of successful scams, the number of neighbors in phishing nodes is generally higher than for normal accounts, making this feature an important marker.

To understand the differences in phishing and non-phishing among the computed features, we compute the mean and standard deviation as shown in table \ref{tab:mean and sd}. Looking the mean, we observe a stark difference in phishing and non-phishing transactions indicating that the problem should be easy, however the standard deviation for both cases is too high indicating that challenge in classification.

\begin{table}
    \centering
    \begin{tabular}{|c| c c c c c c c|}
    \hline
    statistical measure & in-degree & out-degree & degree & in-strength & out-strength & strength & neighbors\\
    \hline
    phishing average & 31.3956 &  20.4905 &  51.8862 &  78.6105 &  86.7360 &  165.3465 & 31.6965\\
    non phishing average & 4.5020 &   4.6438&   9.1459& 72.5328&   9.2551&  81.7880&
    3.6427\\
    phishing std.dev & 180.9905&    96.8388&  219.7376&   691.2912&    860.1017&   1390.3580&
  106.3907\\
  non phishing std.dev & 154.3505&   101.3266&  192.2051& 4409.6850&    281.0234& 4421.5425&
   79.3607\\
    \hline
    \end{tabular}
    \caption{Average and Standard deviation of phishing (1160 nodes) and non-phishing(10800 nodes) nodes feature values}
    \label{tab:mean and sd}
\end{table}

\section{Results and Analysis}
\label{sec:related-and-analysis}

\begin{figure}%
    \centering
    \includegraphics[scale = 0.5]{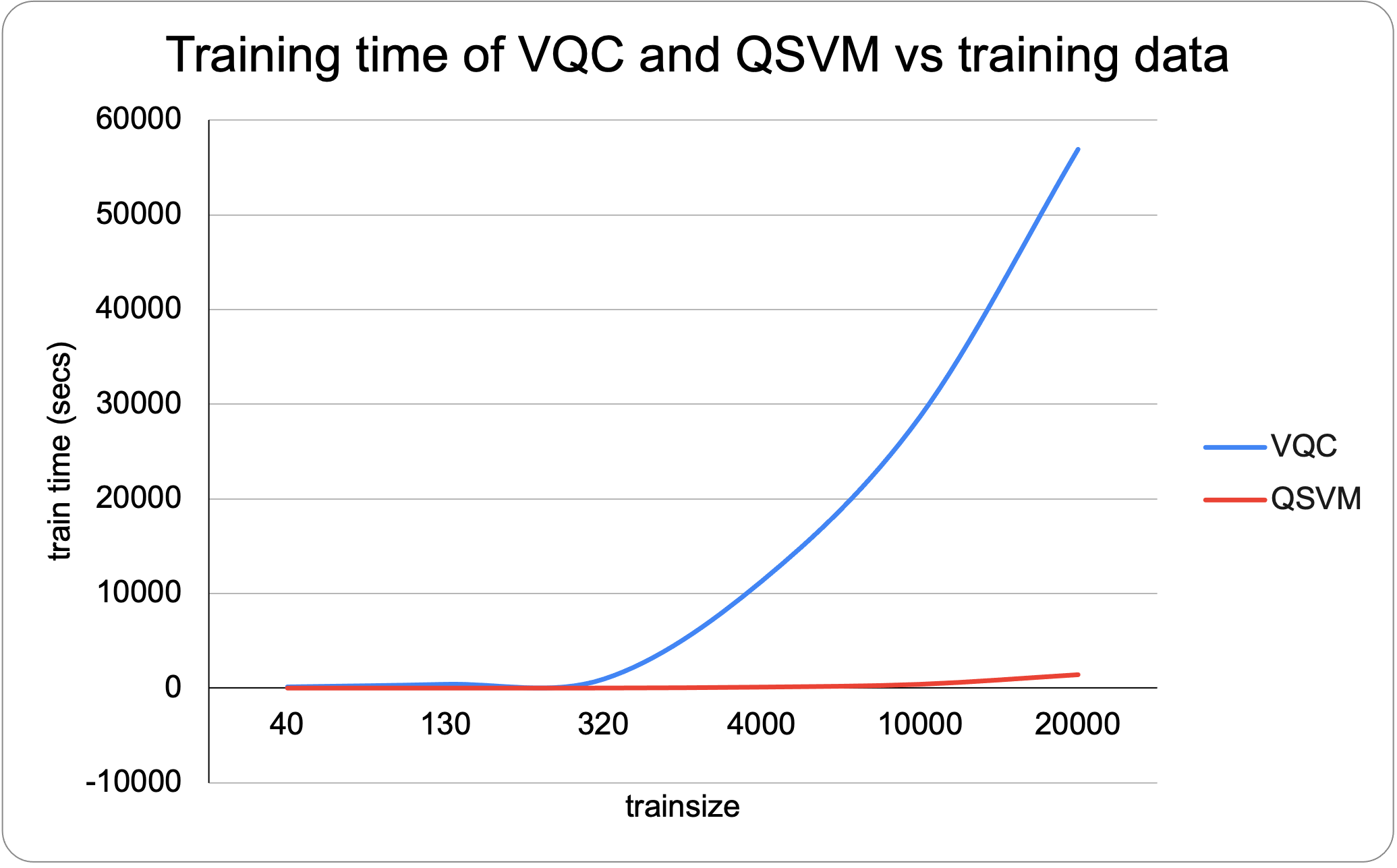}%
    \caption{Training time vs Training data set size for QSVM\_qiskit and VQC models}%
    \label{QSVM-VQC-traintime}%
\end{figure}

This section presents the results using QML, classical ML and the ensemble algorithms on ethereum dataset, ablation studies on QML algorithms and the correlation of ansatz quality metrics with precision, recall and F1-score.
We trained QSVM\_qiskit and VQC on IBM statevector simulator\footnote{IBM Quantum. https://quantum-computing.ibm.com/} using Qiskit \cite{Qiskit} with different training data sizes to understand the scaling of data with respect to computation time. Figure \ref{QSVM-VQC-traintime} shows that the time taken to train QSVM\_qiskit is linearly increasing with increasing train size, whereas train time of VQC increases polynomially with train size. The kernel in QSVM\_qiskit computes an inner product across pairs of data points, thus easier to train than VQC. Figure \ref{QSVM-VQC-traintime} shows that while QSVM\_qiskit finished training with 20K data points in 1389 secs (23 mins), VQC kept optimizing till 56903 secs (15 hours).

\begin{figure}%
    \centering
    \includegraphics[scale = 0.5]{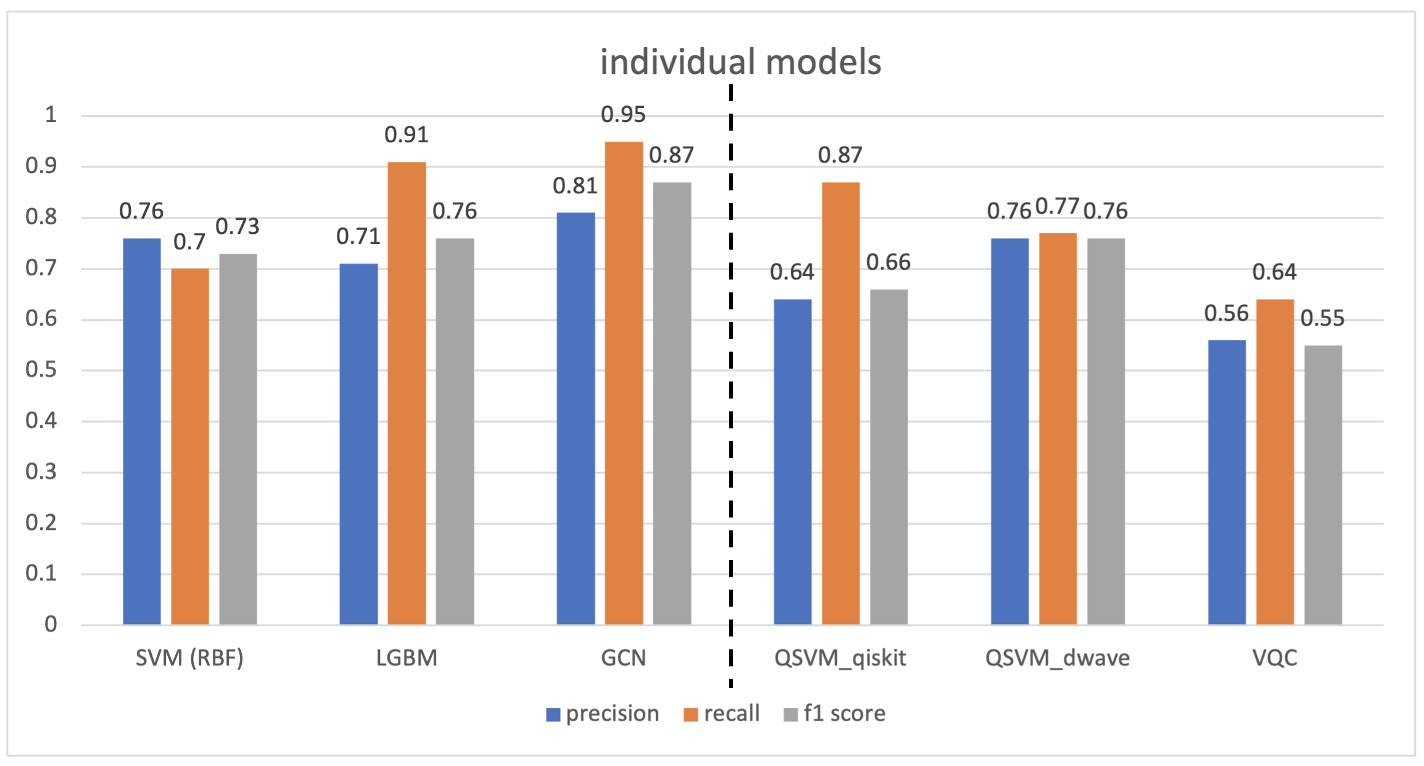}%
    \caption{Macro average of Precision, Recall and F-score of Individual Classical and Quantum model trained using same train (160p-160np) and testdata (1000p-10000np)}%
    \label{fig:indmodel}%
\end{figure}
\begin{figure}%
    \centering
    \includegraphics[scale = 0.45]{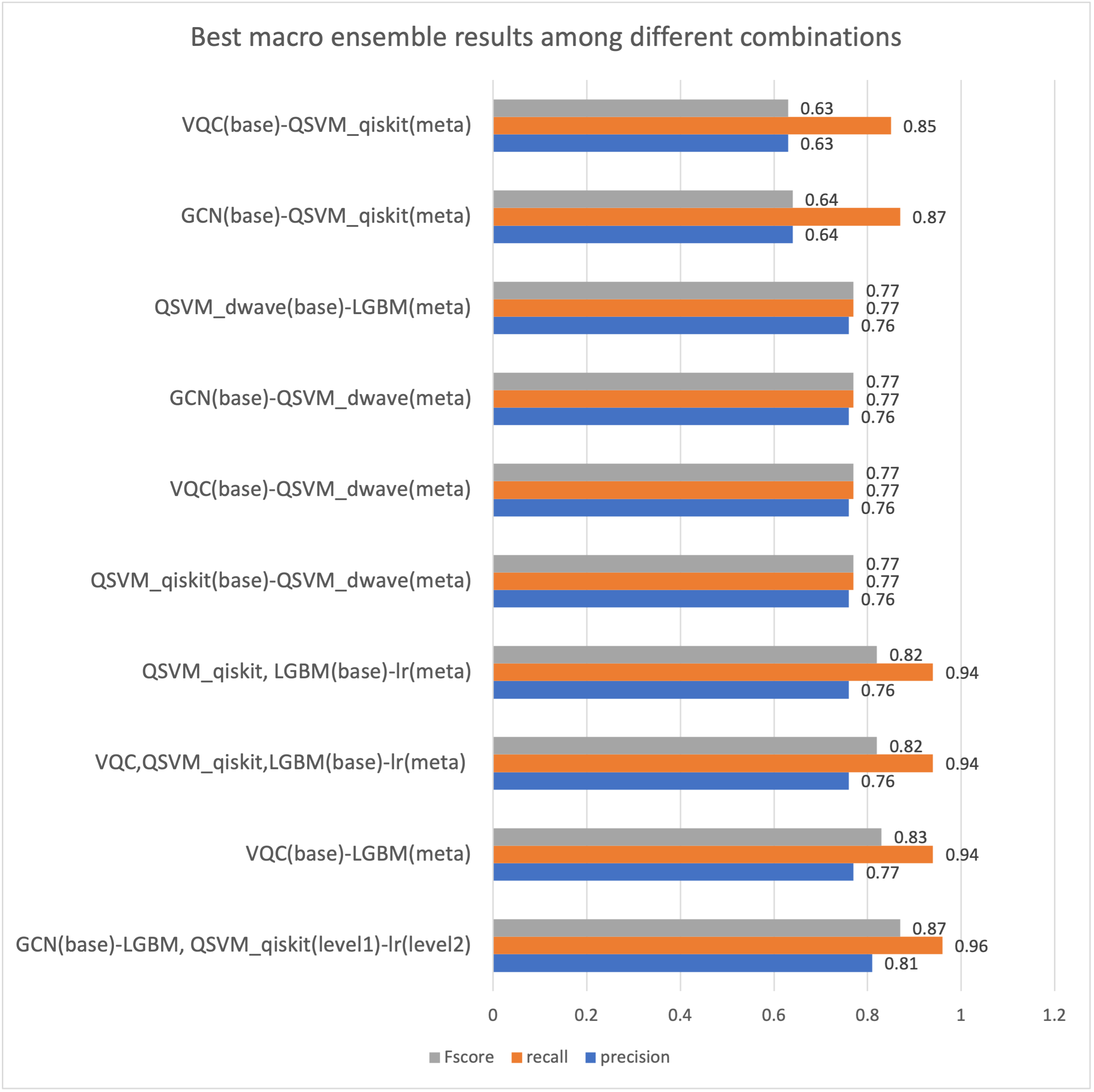}%
    \caption{Macro average of Precision,Recall and f1 scores of best performing Ensembles }%
    \label{fig:best macro}%
\end{figure}

Figure \ref{fig:indmodel} shows the macro average of precision, recall and fscore of each individual quantum and classical algorithms. We observe that classical models are better at recall, while QSVM\_qiskit and VQC are weaker classifiers, thus better candidates for ensembling. QSVM implementation using quantum kernels gave the lowest false positives and best results for phishing class, thus an ideal candidate for our task. 
Figure \ref{fig:best macro} shows the macro average precision, recall, fscore for the top 10 ensemble models and all of them have one or more than one QML algorithm as base or as meta-learners and all using stacking as their strategy. 
Ensemble models with LGBM and a QML algorithm have been able to surpass GCN and in some cases be at par with GCNs metrics while taking significantly less time. QSVM implemented using annealing gave best macro fscore, and was fastest to train in comparison to other quantum models in individual runs or while using bagging, but when stacked with other models it overpowers other models, and the resultant ensembles have same results as that of QSVM annealing run alone, thus obviating an advantage of creating ensemble with this variant of QSVM. Bagging results (shown in appendix) did not have any significant improvement over the individual runs.

\begin{figure}%
    \centering
    \includegraphics[scale = 0.35]{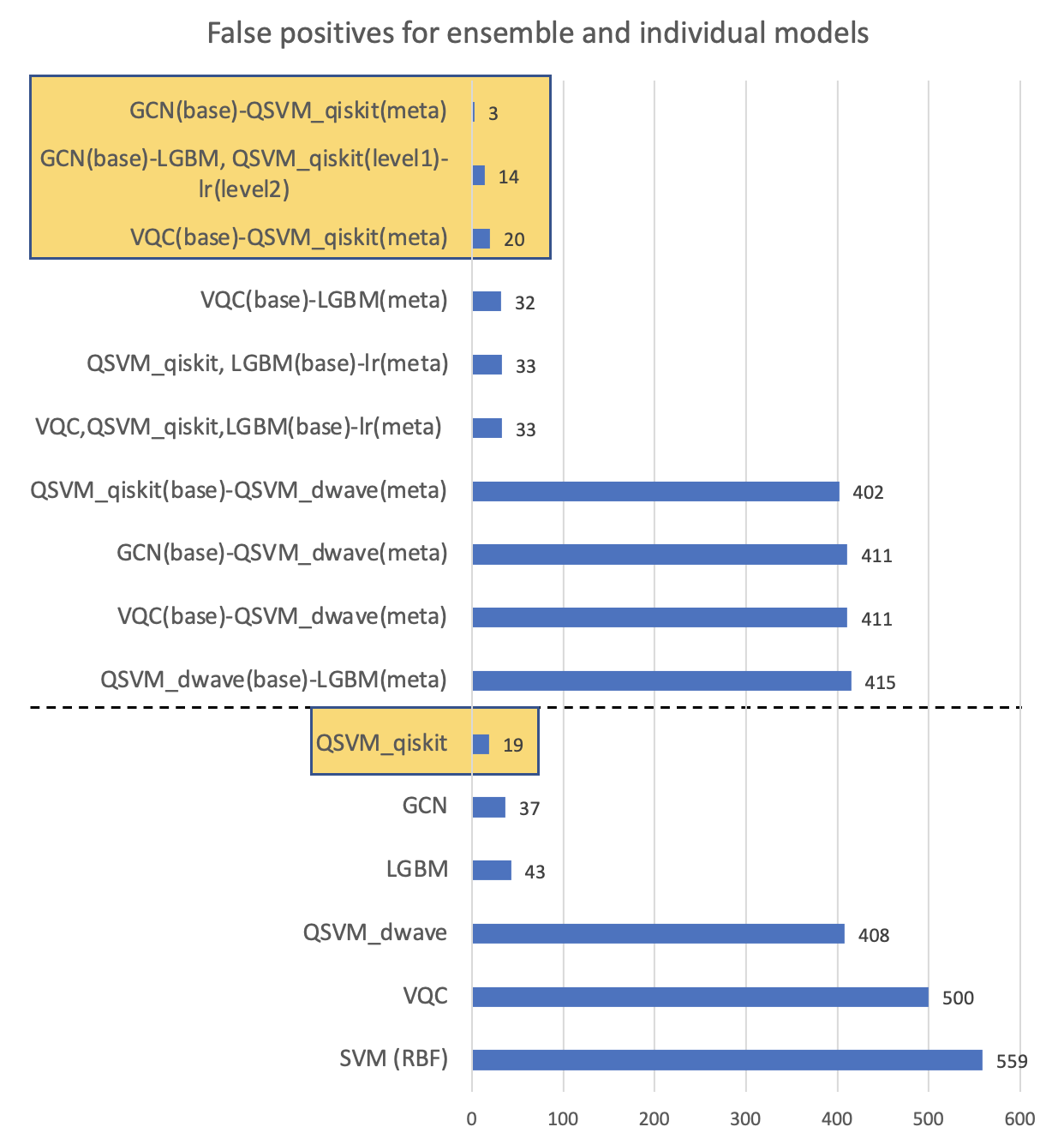}%
    \caption{False positives count of individual models and best performing model combinations }%
    \label{fig:fp}%
\end{figure}

In Table \ref{tab:hardware}, we show that vanilla VQC on a 5 qubit machine performed worse than statevector simulator due to hardware noise. But the important observation is that VQC combined with LGBM helps to improve macro-fscore by 24 points even using the noisy 5 qubit machines. All codes used COBYLA optimizer with maxiter=100. 
\begin{table}
    \centering
    \begin{tabular}{c|c|c|c|c}
    \textbf{Models} & \textbf{Device} & No. of Qubits & \textbf{Macro F1} & \textbf{P\_F1} \\
    \hline
    VQC & statevector & 32 & 0.67 & 0.63\\
    VQC & ibmq\_belem & 5& 0.61 & 0.59 \\
    VQC & ibmq\_geneva & 27 & 0.64 & 0.57\\
    VQC\_lgbm & statevector & 32  & 0.86 & 0.86\\
    VQC\_lgbm & ibmq\_belem & 5 & 0.85 & 0.85\\
    VQC\_lgbm & ibmq\_mumbai & 27 & 0.85 & 0.85\\
    QSVM\_dwave & Classical\_annealing & & 0.78 & 0.74\\
    QSVM\_dwave & D-Wave & 5617 & 0.78 & 0.74\\
    \end{tabular}
    \caption{Results on Quantum Hardware that we had access to and comparing with simulators on same train and test data}
    \label{tab:hardware}
\end{table}

\begin{figure*}%
    \centering
    \includegraphics[width = \textwidth]{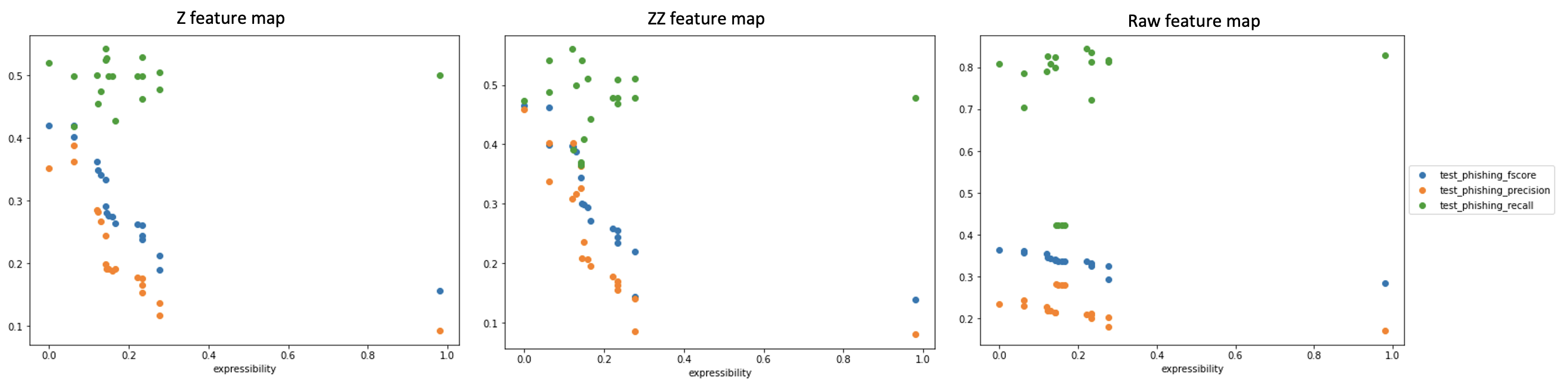}
    \caption{Correlation of Precision (\textbf{orange dots}), Recall (\textbf{green dots}) and F-score (\textbf{blue dots}) with  Expressibility as ansatz quality metric; graph shows correlations for all 19 circuits using Z feature map (left), ZZ feature map(center), and amplitude encoding (right).}
    \label{expressibility}
\end{figure*}

\begin{figure*}[ht]
    \centering
    \includegraphics[width = \textwidth]{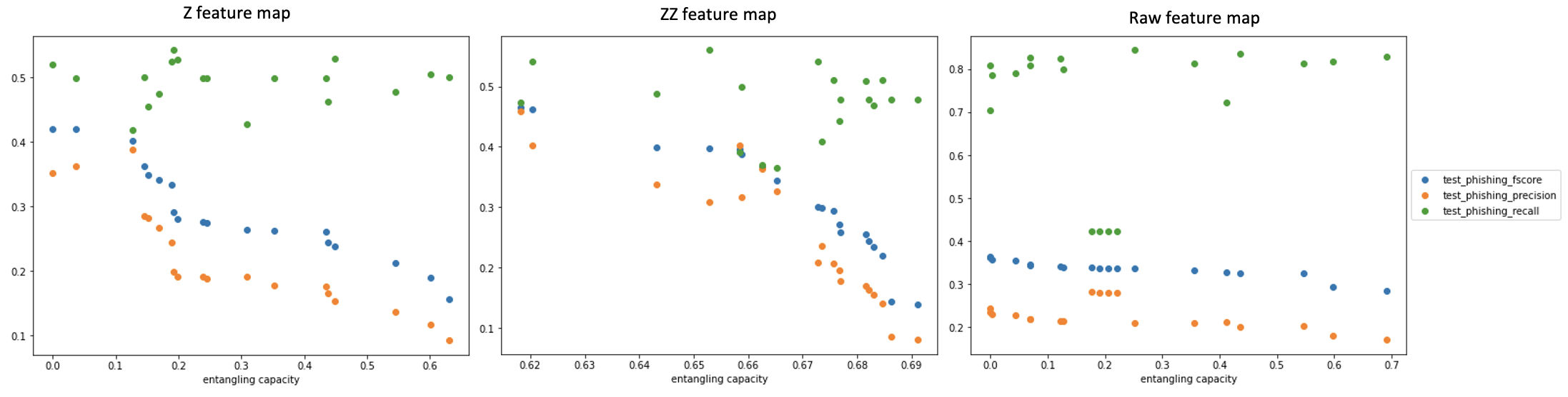}
    \caption{Correlation of Precision (\textbf{orange dots}), Recall (\textbf{green dots}) and F-score (\textbf{blue dots}) with Entangling capacity (x-axis) as ansatz quality metric. Graph shows correlation for all 19 circuits using Z feature map (left), ZZ feature map(center), and amplitude encoding (right).}
    \label{e_capacity}
\end{figure*}

\begin{figure}[ht]
    \centering
    \includegraphics[scale = 0.50]{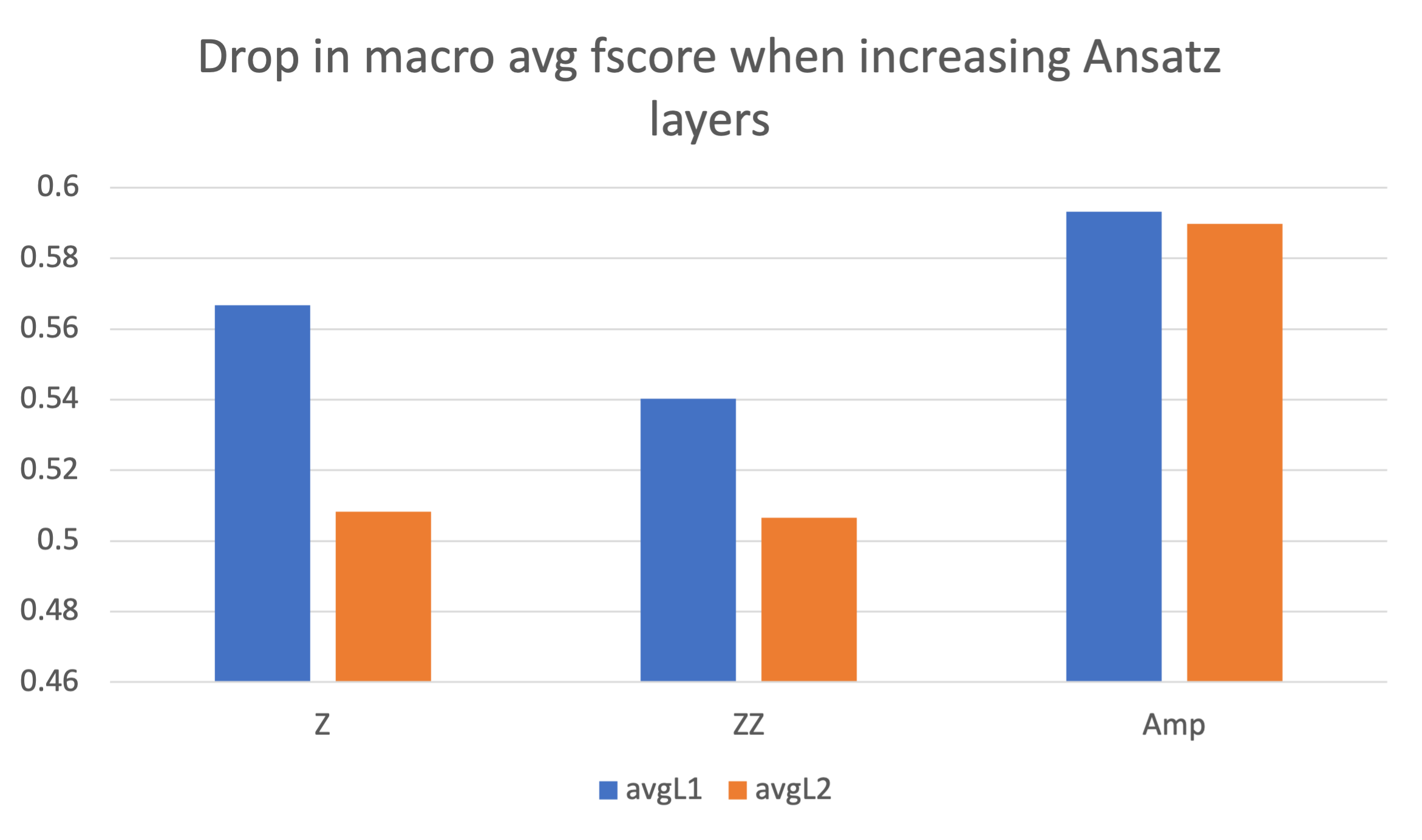}
    \caption{Graph shows that there is a drop in macro average fscore with increasing ansatz layers, and this is true for all three feature maps. Here blue histogram is for 1-layered circuits, and orange is for 2-layered circuits, x-axis has the different feature maps while y-axis is the Macro average Fscore.}
    \label{layers}
\end{figure}
Since we aim to improve phishing detection and ratio of phishing to non-phishing is very low like any other anomaly detection problem, we analyze all models on phishing precision and recall. QSVM\_qiskit has consistently has the lowest number of false positives Figure \ref{fig:fp}, highest number of true negatives, thus when used in ensemble has contributed to increasing precision. This is very important for the application in consideration as we are still okay with some non-phishing transactions being flagged as phishing, while we need to reduce phishing transaction getting mislabeled as non-phishing (false positives). Any combination with QSVM\_qiskit produced more true negatives (least false positives) than other ensembles, except when combined with QSVM\_Dwave, where QSVM\_Dwave overpowered QSVM\_qiskit as shown in figure \ref{fig:fp}.


For VQC, we experiment with 19 benchmarked ansatzes from \cite{Sim_2019}, each with 3 feature maps, done with 1 layer and 2 layers of ansatz, resulting in 114 different experiments and mean of 5 runs is stored. We observed that deeper ansatz with more entangling gates performed well with Z and ZZ features while the same ansatz performs poorly when used with amplitude encoding. In order to be able to analyze the best performing ansatz-feature map pair we explore the correlations between the different input variables. An important observation is shallow anstazes performed better than deeper ansatz, as increasing circuit depth, increases noise and thus we observe a drop in fscore for all 19 circuits from layer 1 to layer 2 as shown in Figure \ref{layers}. This is contrary to deep learning where neural networks benefit from adding more layers as they are able to learn better and dont suffer from decoherence or other quantum phenomenon leading to noise. We intend to study noise mitigation techniques to improve the performance of quantum circuits for these current noisy quantum hardware experiments.  


In order to understand the correlation of ansatz with the problem, we compute expressibility and entangling capacity of each ansatz proposed as ansatz quality metrics in \cite{Sim_2019}. Figure \ref{expressibility} shows correlation of metrics with expressibility for 19 circuits. The key observation is more expressibility doesn't improve precision or fscore for this problem, and as expressibility increases precision and fscore drastically fall. Although for recall, we do not observe any significant change while varying expressibility. The average Pearson correlation coefficient of fscore and expressibility is -0.720 for Z, -0.71295 for ZZ and -0.8413 for amplitude encoding. In case of entangling capacity, we observe a stronger negative correlation with precision and fscore as shown in Figure \ref{e_capacity}. We used Meyer Wallach measure as proposed in \cite{Sim_2019} as well as Von Neumann measure to compute amount of entanglement and both show similar correlation. The average Pearson correlation coefficient of fscore and entangling capacity (using Von Neumann entanglement measure) is -0.940 for Z, -0.919 for ZZ, and -0.940 for amplitude encoding.

\section{Discussions and Future Work}
\label{sec:discussion}
This paper benchmarks Variational Quantum Classifier and Quantum Support vector machine on the problem of phishing detection in Ethereum transaction networks. Detailed ablation studies with varying ansatzes, different feature maps, number of ansatz layers are performed for VQC. We compute expressibility and entangling capacity of each ansatz used and perform correlation studies with these metrics and precision-recall of the phishing detection task. These correlation studies help in choosing the type of ansatz required to improve model performance, instead of heuristic-based ansatzes. For QSVM, we tested the QUBO formulation of QSVM as well as the Quantum Kernel based formulation with the same dataset. For all these models, we ran the best performing hyperparameter set on IBM Quantum devices (5 qubit and 27qubit hardware). In future we aim to develop and use error mitigation techniques to improve the hardware results that are currently getting affected by the noise in the current hardware. In this paper, we also propose an end-to-end trainable framework that creates an ensemble of quantum-classical models and apply it to improve phishing detection in Ethereum network by gaining from the advantages of classical ML models and QML models. Of all the model combinations, we observe a constant decrease in false positives thereby increase in precision in ensembles using QSVM (qiskit). This demonstrates the advantages of quantum kernels and impact of QSVM that can be observed in the individual QSVM runs as well as when used in combination with other models. It is important to understand that current Quantum models are not outperforming their classical counterparts, however can help the classical algorithm in areas where classical algorithm is struggling, for example: improving precision in high class-imbalance scenarios. An interesting future direction is to build quantum ensembles with these models to be able to leverage quantum power in such ensemble models. All quantum circuits and codes are available with tutorials to reproduce the results presented on github: \url{https://github.com/anupamaray/EnsembleQML_application}.


\section*{Acknowledgments}
We are thankful to our colleagues from IBM Research India - Dr Venkat Subramaniam, Dr Yogish Sabharwal and Dr Venkatesan Chakravarthy for giving us the dataset, the financial feature extraction code and all the interesting suggestions that helped us get the best performance in the classical models. The authors thanks Prof Anil Prabhakar (IIT Madras) and Prof Lincy George (IIIT Kotayam) for their valuable suggestions.

\bibliographystyle{unsrt}  
\bibliography{references} 
\end{document}